\documentclass[journal]{IEEEtran}
\usepackage{cite}
\usepackage{csquotes}
\usepackage{graphicx}
\usepackage[outdir=./]{epstopdf}
\usepackage{graphics}
\usepackage{subfigure}
\usepackage{url}
\usepackage{mathtools}
\usepackage{soul}
\usepackage{longtable}
\usepackage{enumitem}
\usepackage{booktabs}
\usepackage{adjustbox}
\usepackage{amssymb}
\usepackage{amsmath}
\usepackage{multirow}
\usepackage{multicol}
\usepackage{caption}
\usepackage{float}
\usepackage{dblfloatfix}
\usepackage{fixltx2e}
\usepackage{afterpage}
\usepackage{verbatim}
\usepackage[justification=centering]{caption}
\usepackage{threeparttable}
\usepackage{hyperref}
\usepackage{color,soul}
\usepackage[many]{tcolorbox}

            \tcbset{
              highlight math style={
                colback=yellow,
                arc=0pt,
                outer arc=0pt,
                boxrule=0pt,
                top=2pt,
                bottom=2pt,
                left=2pt,
                right=2pt,
              }
            }

\begin{document}

\title{CNN-based approach for glaucoma diagnosis using transfer learning and LBP-based data augmentation}
\author{Shishir Maheshwari, Vivek Kanhangad, Ram Bilas Pachori

\thanks{Shishir Maheshwari, Dr. Vivek Kanhangad, and Prof. Ram Bilas Pachori are with Discipline of Electrical Engineering, Indian Institute of Technology Indore, Indore-453552, India (e-mail: phd1501102003@iiti.ac.in, kvivek@iiti.ac.in, pachori@iiti.ac.in).}}

\maketitle

\begin{abstract}
    Glaucoma causes an irreversible damage to retinal nerve fibers which results in vision loss, if undetected in early stage. Therefore, diagnosis of glaucoma in its early stage may prevent further vision loss. In this paper, we propose a convolutional neural network (CNN) based approach for automated glaucoma diagnosis by employing retinal fundus images. This approach employs transfer learning technique and local binary pattern (LBP) based data augmentation. In the proposed approach, we employ Alexnet as a pre-trained CNN model which is used for transfer learning. Initially, the proposed approach divides the fundus image dataset into training and testing data. Further, the color fundus images in training and testing data are separated into red (R), green (G), and blue (B) channels. Additionally, the LBP-based data augmentation is performed on training data. Specifically, we compute LPBs for each of the channel. Finally, the augmented training data is used to train the CNN model via transfer learning. In testing stage, the R, G, and B channels of test image are fed to the trained CNN model which generates $3$ decisions. We employ a decision level fusion technique to combine the decisions obtained from the trained CNN model. The experimental evaluation of the proposed approach on the public RIM-ONE fundus image database, achieves state-of-the-art performance for glaucoma diagnosis. 
\end{abstract}


\section{Introduction}
    As per world health organization (WHO), glaucoma is the second most leading cause for loss of vision after cataract \cite{who}. Glaucoma is also referred to \textit{silent thief of sight} as the symptoms can be noticed only when the disease is quite advanced \cite{silentthief}. In glaucoma, intra ocular pressure (IOP) increases within eye due to the obstruction faced by the flow of aqueous humor. This increased IOP progressively deteriorates optic nerve fibers which is responsible for structural changes in optical nerve head and results in loss of vision. The damages caused by glaucoma are irreversible and cannot be cured permanently but early diagnosis and proper medication may prevent further loss in vision.

    Clinical diagnosis of glaucoma involves various tests suggested by an ophthalmologist at regular intervals. The clinical tests involved are tonometry, gonioscopy, pachymetry, perimetry etc. These tests for clinical diagnosis are manual, time consuming and require skilled supervision. Also, the results are subjective and may vary depending upon intra/inter observer \cite{ura2017}. As the number of people suffering from glaucoma are increasing, relying on the clinical instruments for diagnosing the disease may become unfeasible. Therefore, we need an automated system for glaucoma diagnosis that can aid the experts during mass screening.

    Computer aided diagnosis (CAD) of glaucoma using fundus images can help in automated identification of glaucoma in a fast and accurate manner. CAD systems also manages large number of patients in very less diagnosis time. Of late, various CAD methods for glaucoma have been developed. The major steps involved in typical glaucoma diagnosis approach are: 1. Extraction of discriminating features from fundus images. 2. Classification of the extracted features in order to discriminate between the normal and diseased class. In literature, morphological operations \cite{nayak2009}, wavelet transform \cite{mookiah2012, dua2012, shishir2016,shishir2017}, texture descriptor \cite{shishir2018}, Gabor transform \cite{ura2015_1}, higher order spectra (HOS) \cite{mookiah2012, noronha2014, ura2011} based features are employed. Classifiers such as support vector machine (SVM) \cite{shishir2016,shishir2017,shishir2018}, k-nearest network (KNN) \cite{ura2017}, artificial neural network (ANN) \cite{nayak2009} are used to discriminate the classes. Table \ref{t4-comp} summarizes the existing CAD methodologies developed for glaucoma diagnosis. An efficient combination of feature extraction methods and classifiers is necessary to obtain an effective CAD system. Also, there may arise a possibility that the CAD system developed for a small dataset may not work for large dataset. For a relatively large dataset, it may not show the discriminative ability \cite{uradeep}. Therefore, we need a technique which can extract meaningful features automatically from the data.

    Recently, CNNs have gained popularity among various CAD systems \cite{uradeep}, \cite{deepbrain}, \cite{deepecg}, \cite{deepliver}. CNNs integrates both, an automatic feature extraction and the classification process. CNN architectures perform a series of linear and non-linear operations on the images which extracts meaningful features. CNN is a evolution of multilayer neural network (NN) which involves different layers arranged in different pattern in order to extract the deep features.

    In this paper, we propose a CNN-based approach for automated diagnosis of glaucoma using fundus images. We have utilized transfer learning technique to re-train a CNN model which is designed for other similar classification task. In this approach, we have used Alexnet as a pre-trained CNN model. Specifically, some layers of the pre-trained model are modified in order to deal with binary classification in this work. Moreover, in order to prevent the modified CNN model from over-fitting, the proposed approach employs data augmentation. Specifically, LBP-based data augmentation is employed which computes LBPs from each of the R, G, and B channel.

    The rest of the paper is structured as follow: Section II provides a detailed overview of the proposed approach, CNN architecture, transfer learning, and LBP-based data augmentation. Section III provides details of database and experimental setup. The results and discussion of the proposed approach is presented in section IV. Finally, section V concludes the paper.

                    \begin{figure}[!b]
        				\centering
        				\includegraphics[trim = {5cm 0cm 0cm 0cm},scale=.2]{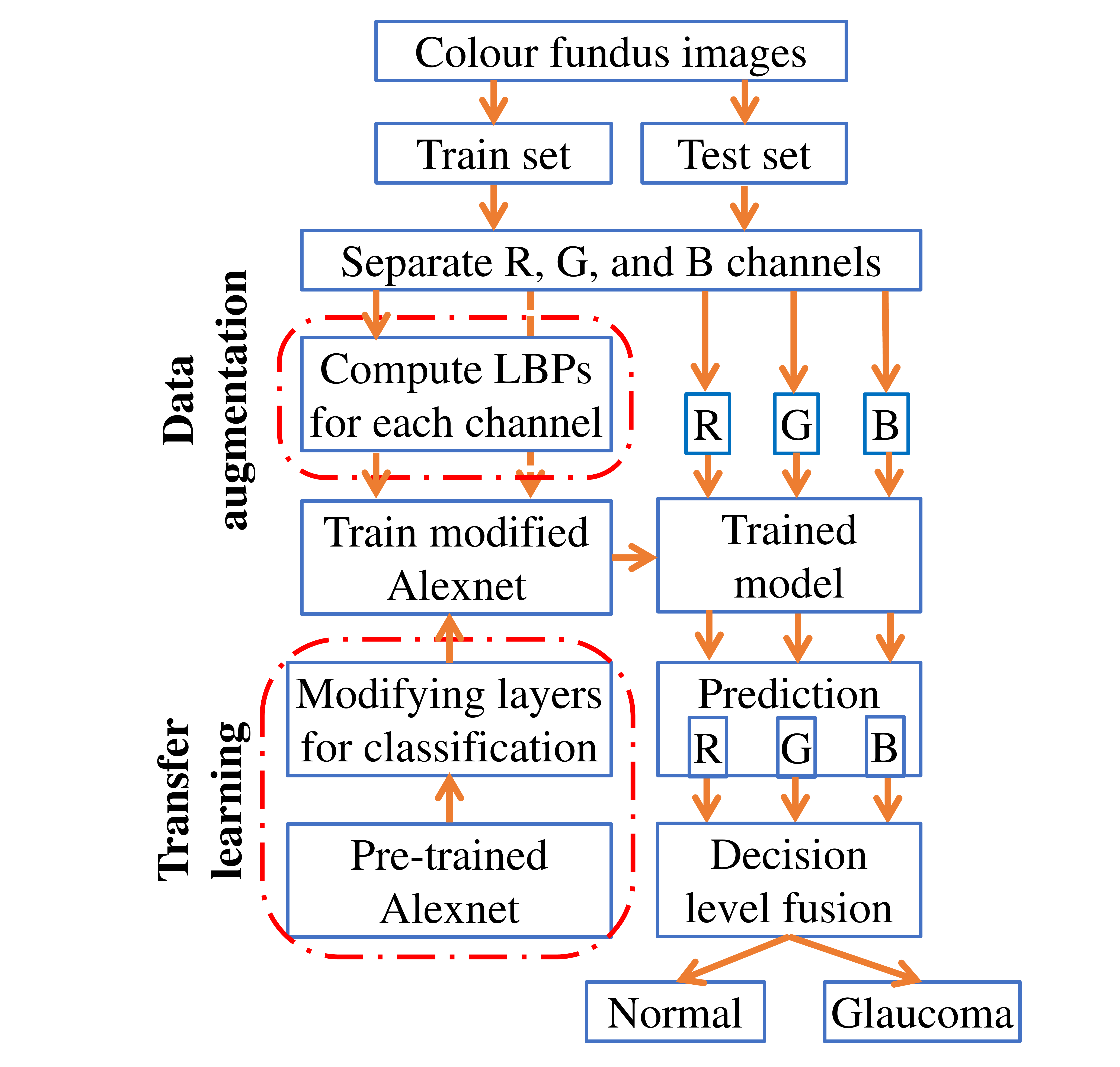}\\
        				\caption{Block diagram of the proposed approach.}\label{f4-blkdia}
        			\end{figure}

\section{CNN based approach for glaucoma diagnosis}
In this section, we briefly explain the steps involved in the proposed approach.
    \subsection{Overview of the proposed approach}
        In this work, a CNN-based approach is proposed for glaucoma diagnosis. The proposed approach employs transfer learning and LBP-based data augmentation. Transfer learning technique is used to re-train a CNN model which is designed for other similar task. Specifically, Alexnet has been employed as a pre-trained CNN model \cite{alexnet}. In order to prevent the model from over-fitting, the data augmentation is employed in this approach. The block diagram of the proposed approach is shown in Fig. \ref{f4-blkdia}. Initially, the fundus image database is randomly split into training and testing data. Further, the colour fundus images in training and testing data are separated into red (R), green (G) and blue (B) channels. Furthermore, the data augmentation process is used to increase the size of training data in order to overcome the problem of over-fitting \cite{alexnet}. Specifically, LBP-based data augmentation is employed by computing LBPs of R, G, and B channels as shown in the data augmentation block in Fig. \ref{f4-blkdia}. Finally, the augmented data is then fed to the modified CNN model for training. During testing stage, the R, G, and B channels of individual test image is fed to the trained model which provides $3$ decisions as shown in Fig. \ref{f4-blkdia}. These decisions are then combined using decision level fusion technique.

                    \begin{table*}[]
				            \centering
				            \caption{Network architecture for modified Alexnet.}\label{t4-modalex}
                            \begin{tabular}{ l c r | l c r }

                            \hline \hline \noalign{\vskip 1mm}
                            \textbf{Layers} & \textbf{Layer label} & \textbf{Layer paramters} & \textbf{Layers} & \textbf{Layer label} & \textbf{Layer paramters} \\
                            \noalign{\vskip 1mm} \hline \noalign{\vskip 1mm}

                            Image Input & data & Sz: 227x227x3 & Convolution & conv5 & Fn: 256, Sz: 3x3x192, St:1, Pd: 1\\
                            \noalign{\vskip 1mm}
                            Convolution  & conv1  & Nf: 96, Sz: 11x11x3, St: 4, Pd=0 & ReLU & relu5 &  - \\
                            \noalign{\vskip 1mm}
                            ReLU & relu1 &  - & Max Pooling & pool5 &  Sz = 3x3, St:2, Pd:0   \\
                            \noalign{\vskip 1mm}
                            Cross Channel Normalization  & norm &  - & Fully Connected     & fc6 &   Sz = 4096\\
                            \noalign{\vskip 1mm}
                            Max Pooling     & pool1 &  Sz: 3x3, St:2 & ReLU  & relu6 &  -\\
                            \noalign{\vskip 1mm}
                            Convolution   & conv2 &   Ft: 256, Sz: 5x5x48, St:1, Pd:2 & Dropout  & drop6 &  50\% \\
                            \noalign{\vskip 1mm}
                            ReLU          &  relu2 &  - & Fully Connected    & fc7 &  Sz: 4096 \\
                            \noalign{\vskip 1mm}
                            Cross Channel Normalization  & norm2 &  - & ReLU   & relu7 & -\\
                            \noalign{\vskip 1mm}
                            Max Pooling    & pool2 &  Sz: 3x3, St:2 & Dropout  & drop7 &    50\% \\
                            \noalign{\vskip 1mm}
                            Convolution & conv3 & Nf: 384. Sz: 3x3x256, St: 1, pd:1 &  Fully Connected & fc8 &  Sz: 2 \\
                            \noalign{\vskip 1mm}
                            ReLU            & relu3 & - & Softmax  & prob &   -  \\
                            \noalign{\vskip 1mm}
                            Convolution & conv4 & Fn: 384, Sz: 3x3x192, St:1 Pd:1 & Classification Output & output & cross-entropy \\
                            \noalign{\vskip 1mm}
                            ReLU & relu4 &  - & & & \\
                            \noalign{\vskip 1mm} \hline \noalign{\vskip 1mm}

                        \end{tabular} 
                        \begin{tablenotes}[flushleft]
                            \small
                            \item *Fn = number of filters, Sz = filter size, Pd = padding, St = Stride
                        \end{tablenotes}
                    \end{table*}

    \subsection{Convolutional neural network}
        CNN is an advance neural network (NN) architecture which finds its application in many computer vision applications such as face recognition \cite{deepface}, segmentation \cite{deepseg}, and other classification based approaches \cite{deepbrain} \cite{deepliver}. Traditional methods for glaucoma diagnosis involve two step process: feature extraction followed by classification. Whereas, CNN is one step process which integrates both, the feature extraction and classification.

        NN is inspired by human brain and is used in machine learning. Multi-layer perceptron (MLP) is a basic NN composed of input layer, hidden layer/layer's, and output layer. The hidden layer consists of one or more layers of neurons connected with previous layer. These neurons learns the feature from input data by adjusting the weights. When an unexpected output is obtained for a given set of input, the weights are adjusted by back propagation algorithm to obtain the desired output \cite{allan}.

        CNN is a multi-layered neural network which is suitable for input data of high-dimension such as images as it can achieve some kind of shift and deformation invariance \cite{cnn}. CNNs are also composed of neurons, but are arranged in 3D pattern having width, height and number of filters. The CNN architecture is composed of one or more repetition of following layers:
             \begin{enumerate}
                \item Input layer: This is the first layer of CNN which holds and passes the raw data to other connected layers for further processing.
                \item Convolutional layer: This layer performs a dot product between filter and local image region of size equal to filter. It is referred to as learning layer.
                \item Rectified linear unit (ReLU): This is thresh-holding layer which applies a activation function of max($0$, $x$). If $x$ is less/more than $0$, the output will be $0$/$x$, respectively.
                \item Max-pooling layer: It reduces the spatial dimension of the data received from the previous layer.
                \item Fully connected layer: In this layer, the neurons are fully connected with the neurons of previous layer.
                \item Soft-max layer: This layer brings the data in range $0$ - $1$. It is a normalized exponential function.
                \item Output layer: This layer provides the output prediction. It contains the loss function and label of the input data.
             \end{enumerate}

                            \begin{figure*}
                              \centering
                              \includegraphics[scale=.10]{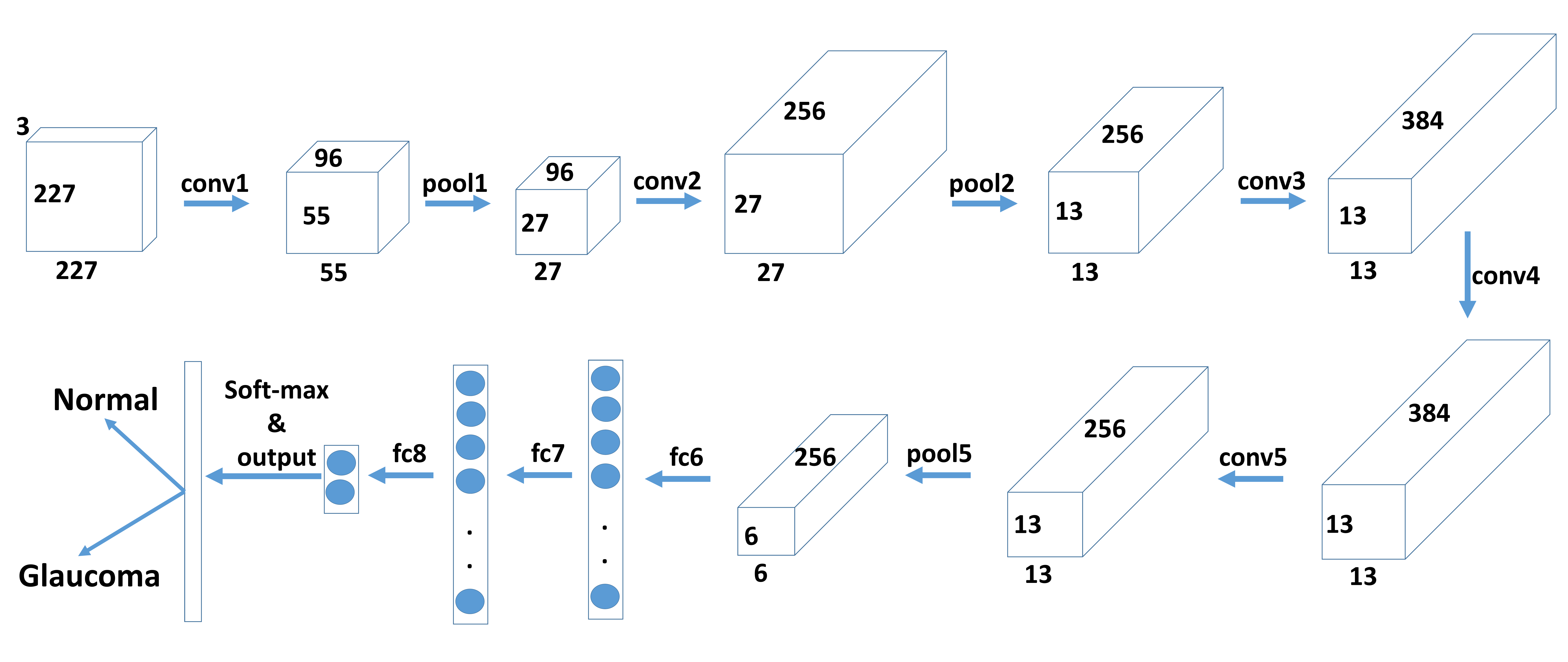}\\
                              \caption{Modified Alexnet architecture via transfer learning.}\label{f4-cnn}
                            \end{figure*}

    \subsection{Transfer learning}
        Training an entire CNN model from scratch requires a very large labelled dataset and a capable GPU-based hardware. It is not feasible to train a network from scratch due to hardware limitation and small size of the dataset. To overcome this limitation, transfer learning technique is employed in this work. Transfer learning is a technique which uses a pre-trained CNN model already trained for other similar task. In this technique, a pre-trained CNN model is trained on new data for similar classification task by changing certain design parameters \cite{translearn}.

        Training a pre-trained CNN model is about adjusting the layer weights according to new training data. In the proposed approach, we have employed a pre-trained Alexnet model \cite{alexnet} which needs to be trained on new data via transfer learning. Alexnet contains eight layers, out of which five are convolution layers and three are fully-connected layers. Originally, Alexnet is capable of classifying between $1000$ classes. However, in the proposed approach, Alexnet is modified to perform binary classification. Specifically, the last three layers namely fully-connected layer, soft-max layer and the classification layer are modified and trained from scratch. Table \ref{t4-modalex} lists the different layers and their dimensions of modified Alexnet. Fig. \ref{f4-cnn} shows the modified Alexnet architecture used in this work.

    \subsection{Data augmentation}
        In order to prevent the CNN model from over-fitting, a large number of labelled data is required to train the CNN model \cite{alexnet}. As the size of the training data available for this task is small, data augmentation method is employed to increase the size of training data. Data augmentation employ operations such as cropping, scaling, translation, shear, zooming, rotation, and reflection. However, in the proposed approach, we have not used these operation for data augmentation. Instead, we have computed LBPs \cite{lbp2002} of R, G, and B channels of the training data. The LBP is an effective gray level texture descriptor \cite{lbp2002} which has also been used for face verification \cite{lbpface}. The LBP-based data augmentation block is shown in Fig. \ref{f4-blkdia}. The computed LBPs of each channels are 2D vector, so we read it as LBP images.

                \begin{table*}[!t]
				\centering
				\caption{Classification performance for LBP-based data augmentation.}\label{t4-explbp}
				\resizebox{1.8\columnwidth}{!}{$
				\begin{tabular}{l c c c r}
					\hline \hline \noalign{\vskip 1mm}					
					\textbf{Data split} & \textbf{Performance of} & \textbf{Accuracy} (\%)& \textbf{Sensitivity} (\%) &
                    \textbf{Specificity} (\%) \\
                     (training:testing)& & (max-mean-min) & (max-mean-min) & (max-mean-min) \\
					\noalign{\vskip 1mm}  \hline \noalign{\vskip 1mm}
					\noalign{\vskip 1mm}

					$90:10$ & R channel & 98.90 - 94.28 - 92.30 & 100 - 97.25 - 90.19 & 97.50 - 90.50 - 82.50 \\
                    & G channel & 94.50 - 91.37 - 90.11 & 100 - 96.27 - 90.19 & 97.50 - 86.00 - 75.00 \\
                    & B channel & 93.40 - 91.42 - 90.11 & 98.03 - 94.80 - 92.15 & 92.50 - 87.12 - 82.50  \\
                    & Decision level fusion & 96.70 - 93.40 - 92.30 & 100 - 96.69 - 92.15 & 97.50 - 88.75 - 85.00 \\
					\noalign{\vskip 1mm} \hline \noalign{\vskip 1mm}

                    $80:20$ & R channel & 95.58 - 92.86 - 91.91 & 100 - 95.68 - 90.79 & 95.00 - 89.33 - 86.67 \\
                    & G channel & 92.65 - 90.69 - 88.97 & 100 - 94.80 - 89.47 & 95.00 - 85.67 - 75.00 \\
                    & B channel & 92.65 - 89.55 - 88.33 & 98.68 - 93.22 - 86.84 & 90.0 - 84.33 - 76.67  \\
                    & Decision level fusion & 94.11 - 92.24 - 91.17 & 100 - 95.26 - 88.15 & 96.67 - 88.41 - 81.67 \\
					\noalign{\vskip 1mm} \hline \noalign{\vskip 1mm}

                    $70:30$ & R channel & 99.10 - 96.00 - 93.33 & 100 - 97.20 - 92.00 & 100 - 94.50 - 90.00 \\
                    & G channel & 98.95 - 95.00 - 93.33 & 100 - 98.60 - 92.00 & 100 - 89.00 - 70.00 \\
                    & B channel & 98.60 - 94.33 - 93.33 & 100 - 96.20 - 92.00 & 100 - 92.00 - 85.00  \\
                    & Decision level fusion & 100 - 96.00 - 93.33 & 100 - 98.00 - 92.00 & 100 - 93.50 - 85.00 \\
					\noalign{\vskip 1mm} \hline \hline
				\end{tabular}
				$}
			\end{table*}

\section{Dataset \& experimental setup}
    \subsection{Dataset}
         We have used publicly available RIM-ONE database in this work. It contains $455$ color fundus images with $255$ normal and $250$ glaucoma images. This database is obtained from medical image analysis group (MIAG) and is available at \href{http://medimrg.webs.ull.es/}{http://medimrg.webs.ull.es/}. The fundus images in the dataset are stored in JPEG format with different resolutions.

    \subsection{Experimental setup}\label{l4-expsetup}
        The fundus image database is divided into training ($80$\%) and testing ($20$\%) data. We have also performed our experiments with $70:30$ and $90:10$ data split into training:testing. The images in training and testing data is separated into R, G, and B channels. Further, LBP based data augmentation is employed by computing LBPs of individual channels of training data. We have not performed LBP based data augmentation on testing data. In order to train the model, values of the parameters involved are initialized as follows: learning rate is set to $0.0001$, stochastic gradient descent with momentum (SGDM) optimizer is used as a solver, number of images in a batch is $20$, and number of epochs are $80$.

\section{Results \& discussion}
    \subsection{Results}
        In order to access the performance of the proposed approach, the performance metric parameters such as accuracy, sensitivity and specificity \cite{azar2014} are employed. The experiments are repeated $20$ times and the classification performances are tabulated in the form of $maximum-mean-minimum$ value in Table \ref{t4-explbp}.

        Table \ref{t4-explbp} tabulates classification performances for $70$:$30$, $80$:$20$ and  $90$:$10$ training:testing data split. It can be observed from the Table \ref{t4-explbp} that the R channel individually achieves maximum classification accuracy of 99.10\%, 98.90\% and 95.58\% for $70$:$30$, $80$:$20$ and $90$:$10$ training:testing data split, respectively. The B channel individually obtains minimum performance. We have also performed experiments without employing LBP-based data augmentation for $80$:$20$ training:testing data split. The classification performance without LBP-based data augmentation is tabulated in Table \ref{t4-expnolbp}. It can be observed from Table \ref{t4-expnolbp} that the experiments performed with LBP-based data augmentation achieves better classification performance when compared to the experiments performed without employing LBP-based data augmentation.

                            \begin{figure*}
                              \centering
                              \includegraphics[scale=.5]{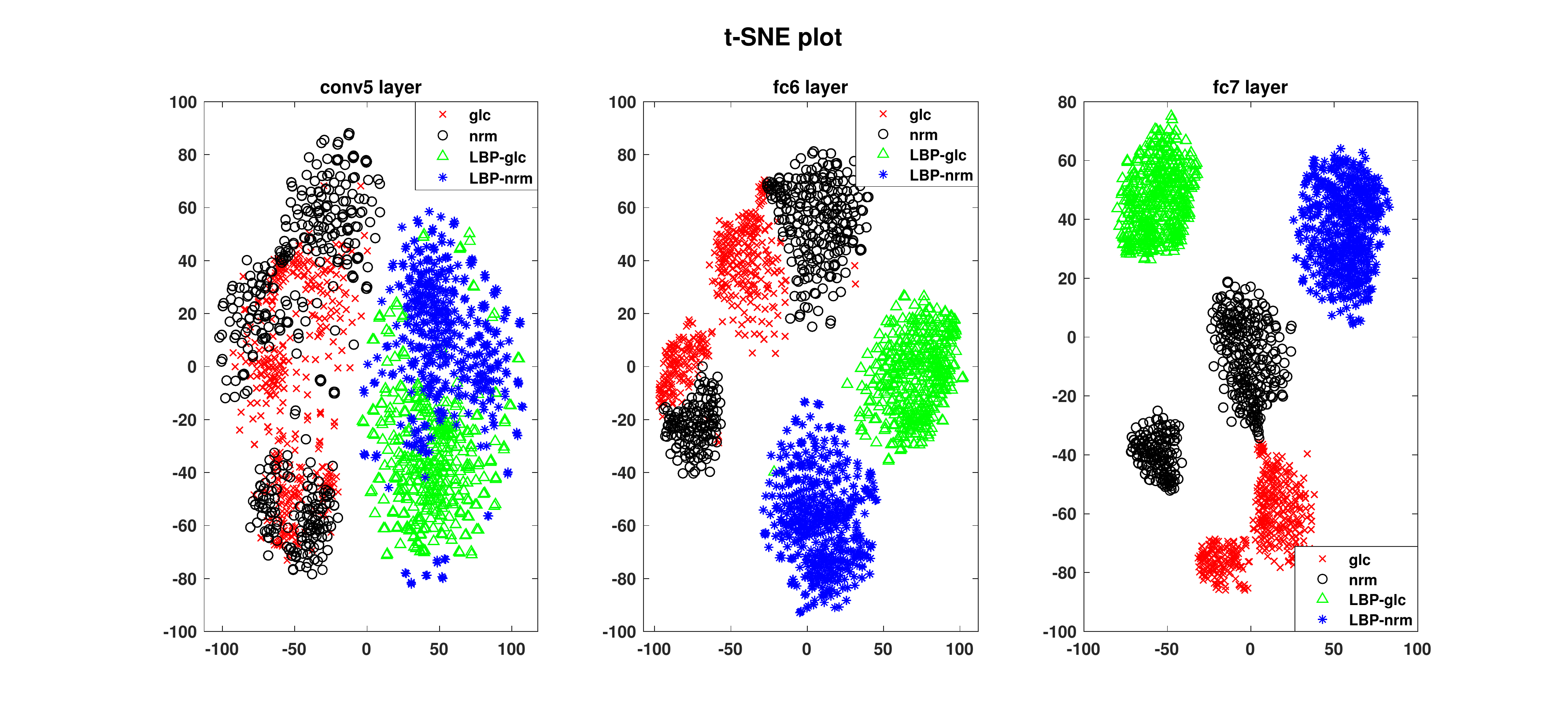}
                              \caption{The t-SNE plot of: features learned by conv5 layer (left), output of fc6 (middle) and fc7 (right) layers of the modified CNN model.}\label{f4-tsne}
                            \end{figure*}

            \begin{table}[]
				\centering
				\caption{Classification performance without LBP-based data augmentation for $80$:$20$
                    training:testing data split.}\label{t4-expnolbp}
				\resizebox{.95\columnwidth}{!}{$
				\begin{tabular}{l c r}
					\hline \hline \noalign{\vskip 1mm}					
					\textbf{LBP-based} & \textbf{Channel} & \textbf{Accuracy} (\%)\\
                    \textbf{data augmentation} & & (max-mean-min) \\
					\noalign{\vskip 1mm}  \hline \noalign{\vskip 1mm}
					\noalign{\vskip 1mm}
					Yes & R & 98.90 - 94.28 - 92.30 \\
                        & G & 94.50 - 91.37 - 90.11 \\
                        & B & 93.40 - 91.42 - 90.11 \\
                    \noalign{\vskip 1mm}  \hline \noalign{\vskip 1mm}
                    No  & R & 94.51 - 92.97 - 92.31 \\
                        & G & 93.41 - 91.10 - 90.11 \\
                        & B & 93.41 - 90.44 - 89.01 \\
					\noalign{\vskip 1mm} \hline \hline
				\end{tabular}
				$}
			\end{table}

        \begin{table*}[]
				\centering
                \caption{A comparative summary of the existing methods for automated glaucoma diagnosis for public database.}\label{t4-comp}
				\resizebox{1.7\columnwidth}{!}{$
				\begin{tabular}{ l c c r }
				\hline \hline \noalign{\vskip 1mm}

                \textbf{Papers} & \textbf{Method} & \textbf{Classifier/CNN architecture} & \textbf{Performance parameters} (\%) \\
				
                \noalign{\vskip 1mm} \hline \noalign{\vskip 1mm}
				\noalign{\vskip 1mm}					

                    \cite{ura20171} & Non parametric spatial &  SVM & ac: 93.62 \\
					& envelop energy pattern & & sn: 87.50 \\
					& & & sp: 98.43 \\
					\noalign{\vskip 1mm} \hline \noalign{\vskip 1mm}

					\cite{bander} & Deep CNN &  SVM & ac: 88.20 \\
					& based feature extraction & & sn: 90.8 \\
					& & & sp: 85 \\
					\noalign{\vskip 1mm} \hline \noalign{\vskip 1mm}

					\cite{allan} & Deep CNN &  Google-net & ac: 86.20 \\
					& & & sn: - \\
					& & & sp: - \\
					\noalign{\vskip 1mm} \hline \noalign{\vskip 1mm}

					\cite{shishir2016} & Empirical wavelet transform &  LS-SVM & ac: 81.32 \\
					& based correntropy features & & sn: - \\
					& & & sp: - \\
					\noalign{\vskip 1mm} \hline \noalign{\vskip 1mm}

					\cite{kirar1} & Discrete \& empirical &  LS-SVM & ac: 83.57 \\
					& wavelet transform  & & sn: 86.40 \\
					& & & sp: 80.80 \\
					\noalign{\vskip 1mm} \hline \noalign{\vskip 1mm}

					\cite{shishir2017} & Variational mode decomposition &  SVM & ac: 81.62 \\
					& based entropy and fractal features & & sn: - \\
					& & & sp: - \\
					\noalign{\vskip 1mm} \hline \noalign{\vskip 1mm}

                    \textbf{Proposed method} & \textbf{Local binary pattern} & \textbf{Alexnet} & \textbf{ac: 98.90}\\
					&  \textbf{\& Deep CNN transfer learning}  & & \textbf{sn: 100}\\
					& & & \textbf{sp: 97.50} \\
					
					\noalign{\vskip 1mm}  \hline
					\hline    \noalign{\vskip 1mm}

				\end{tabular} $}\\
					\begin{tablenotes}[flushleft]
                    \small
                    \item *NR = Not reported, ac = accuracy, sn = sensitivity, sp = specificity, AROC = Area under receiver operating characteristics
                    \end{tablenotes}
			\end{table*}

    \subsection{Discussion}
        Table \ref{t4-comp} presents the brief description of past methodologies developed for automated glaucoma diagnosis using public database. Traditional methodologies for diagnose systems typically involves: feature extraction and classification. Though, the selection of relevant combination of feature set and classifier is tedious. However, deep CNN networks automatically extracts relevant features by its own and classifies them into different classes. The advantage of CNN based methods is that one need not perform explicit feature extraction, statistical analysis, ranking and classification. The proposed approach is based on transfer learning of Alexnet which overcomes the limitation of advance hardware and the time require to train the network from scratch.

        In order to prevent the model from over-fitting, data augmentation method is employed which increases size of the training data. LBP-based data augmentation has been in this work. The data is augmented by computing LBPs of each of the channel. The LBP based data augmentation block is shown in Fig. \ref{f4-blkdia}. LBP is simple and effective feature descriptor which is employed for texture based image classification in \cite{lbptext1}, \cite{lbptext2}. In the proposed approach, to demonstrate the effectiveness of LBP-based data augmentation, we have employed t-Distributed Stochastic Neighbor Embedding (t-SNE) \cite{tsne} method to visualize the non-LBP and LBP features learned by the CNN model. The t-SNE is a nonlinear dimensionality reduction technique which is well-suited for embedding high-dimensional data for visualization in a low-dimensional space of two or three dimensions \cite{tsne}. Fig. \ref{f4-tsne} shows the t-SNE plot for the high dimensional LBP and non-LBP features learned by \textit{conv5}, \textit{fc6}, and \textit{fc7} layers of the CNN model. It can be observed from the Fig. \ref{f4-tsne} that LBP features learned by CNN layers show better inter-class discrimination as compared to non-LBP features.

        We have also performed the experiments without employing LBP-based data augmentation for training data. The result of which is tabulated in Table \ref{t4-expnolbp}. It can be observed from Table \ref{t4-expnolbp} and Fig. \ref{f4-tsne} that due to good inter-class discrimination capability of LBP features, the classification performance of the CNN model improves by employing LBP-based data augmentation.

        The advantages of the proposed approach are:
        \begin{enumerate}
                \item It can be employed for pre-screening of fundus images to aid the clinicians during mass-screening as it achieves promising classification performance.
                \item It involves transfer learning of pre-trained CNN network which helps in overcoming the limitation of capable hardware and potential reduction in time for designing a network from scratch.
        \end{enumerate}

        The limitation of the proposed approach is that it is evaluated on small dataset. In future, the authors propose to test the proposed approach on large database to evaluate its effectiveness. Also, proper selection of training parameters may help in improving the performance of the approach.

\section{Conclusion}
    In this paper, we have presented a CNN based approach for automated glaucoma diagnosis. The proposed approach employs transfer learning technique to re-train the already trained CNN model for some other similar task. Transfer learning is employed to overcome the limitation of having advance and GPU-based capable hardware. To prevent the CNN model from over-fitting, LBP-based data augmentation technique is employed which increases the size of training data. The experimental results suggests that the performance of the proposed approach improves by employing LBP-based data augmentation. The approach obtains promising classification performance which can help in reducing the burden on experts during mass screening. The authors, in future, intend to extend the proposed work on other retinal diseases.

\bibliographystyle{IEEEtran}
\bibliography{glaucoma_class_bibfile.bbl}

\begin{thebibliography}{10}
\providecommand{\url}[1]{#1}
\csname url@samestyle\endcsname
\providecommand{\newblock}{\relax}
\providecommand{\bibinfo}[2]{#2}
\providecommand{\BIBentrySTDinterwordspacing}{\spaceskip=0pt\relax}
\providecommand{\BIBentryALTinterwordstretchfactor}{4}
\providecommand{\BIBentryALTinterwordspacing}{\spaceskip=\fontdimen2\font plus
\BIBentryALTinterwordstretchfactor\fontdimen3\font minus
  \fontdimen4\font\relax}
\providecommand{\BIBforeignlanguage}[2]{{%
\expandafter\ifx\csname l@#1\endcsname\relax
\typeout{** WARNING: IEEEtran.bst: No hyphenation pattern has been}%
\typeout{** loaded for the language `#1'. Using the pattern for}%
\typeout{** the default language instead.}%
\else
\language=\csname l@#1\endcsname
\fi
#2}}
\providecommand{\BIBdecl}{\relax}
\BIBdecl

\bibitem{who}
H.~A. Quigley and A.~T. Broman, ``The number of people with glaucoma worldwide
  in 2010 and 2020,'' \emph{British Journal of Ophthalmology}, vol.~90, no.~3,
  pp. 262--267, 2006.

\bibitem{silentthief}
X.~Chen, Y.~Xu, S.~Yan, D.~W.~K. Wong, T.~Y. Wong, and J.~Liu, ``Automatic
  feature learning for glaucoma detection based on deep learning.''\hskip 1em
  plus 0.5em minus 0.4em\relax Springer International Publishing, 2015, pp.
  669--677.

\bibitem{ura2017}
U.~R. Acharya, S.~Bhat, J.~E. Koh, S.~V. Bhandary, and H.~Adeli, ``A novel
  algorithm to detect glaucoma risk using texton and local configuration
  pattern features extracted from fundus images,'' \emph{Computers in Biology
  and Medicine}, vol.~88, pp. 72 -- 83, 2017.

\bibitem{nayak2009}
J.~Nayak, U.~R. Acharya, P.~S. Bhat, N.~Shetty, and T.~C. Lim, ``{Automated
  diagnosis of glaucoma using digital fundus images},'' \emph{Journal of
  Medical Systems}, vol.~33, no.~5, pp. 337--346, 2009.

\bibitem{mookiah2012}
M.~R.~K. Mookiah, U.~R. Acharya, C.~M. Lim, A.~Petznick, and J.~S. Suri,
  ``{Data mining technique for automated diagnosis of glaucoma using higher
  order spectra and wavelet energy features},'' \emph{Knowledge-Based Systems},
  vol.~33, pp. 73--82, 2012.

\bibitem{dua2012}
S.~Dua, U.~R. Acharya, P.~Chowriappa, and S.~V. Sree, ``{Wavelet based energy
  features for glaucomatous image classification},'' \emph{IEEE Transactions on
  Information Technology in Biomedicine}, vol.~16, no.~1, pp. 80--87, 2012.

\bibitem{shishir2016}
S.~Maheshwari, R.~B. Pachori, and U.~R. Acharya, ``Automated diagnosis of
  glaucoma using empirical wavelet transform and correntropy features extracted
  from fundus images,'' \emph{IEEE Journal of Biomedical and Health
  Informatics}, vol.~21, no.~3, pp. 803--813, May 2017.

\bibitem{shishir2017}
S.~Maheshwari, R.~B. Pachori, V.~Kanhangad, S.~V. Bhandary, and U.~R. Acharya,
  ``Iterative variational mode decomposition based automated detection of
  glaucoma using fundus images,'' \emph{Computers in Biology and Medicine},
  vol.~88, pp. 142 -- 149, 2017.

\bibitem{shishir2018}
S.~Maheshwari, V.~Kanhangad, R.~B. Pachori, S.~V. Bhandary, and U.~R. Acharya,
  ``Automated glaucoma diagnosis using bit-plane slicing and local binary
  pattern techniques,'' \emph{Computers in Biology and Medicine}, vol. 105, pp.
  72 -- 80, 2019.

\bibitem{ura2015_1}
U.~R. Acharya, E.~Ng, L.~W.~J. Eugene, K.~P. Noronha, L.~C. Min, K.~P. Nayak,
  and S.~V. Bhandary, ``{Decision support system for the glaucoma using Gabor
  transformation},'' \emph{Biomedical Signal Processing and Control}, vol.~15,
  pp. 18--26, 2015.

\bibitem{noronha2014}
K.~P. Noronha, U.~R. Acharya, K.~P. Nayak, R.~J. Martis, and S.~V. Bhandary,
  ``{Automated classification of glaucoma stages using higher order cumulant
  features},'' \emph{Biomedical Signal Processing and Control}, vol.~10, pp.
  174--183, 2014.

\bibitem{ura2011}
U.~R. Acharya, S.~Dua, X.~Du, S.~V. Sree, and C.~K. Chua, ``{Automated
  diagnosis of glaucoma using texture and higher order spectra features},''
  \emph{IEEE Transactions on Information Technology in Biomedicine}, vol.~15,
  no.~3, pp. 449--455, 2011.

\bibitem{uradeep}
U.~Raghavendra, H.~Fujita, S.~V. Bhandary, A.~Gudigar, J.~H. Tan, and U.~R.
  Acharya, ``Deep convolution neural network for accurate diagnosis of glaucoma
  using digital fundus images,'' \emph{Information Sciences}, vol. 441, pp. 41
  -- 49, 2018.

\bibitem{deepbrain}
M.~Talo, U.~B. Baloglu, Özal Yıldırım, and U.~R. Acharya, ``Application of
  deep transfer learning for automated brain abnormality classification using
  mr images,'' \emph{Cognitive Systems Research}, vol.~54, pp. 176 -- 188,
  2019.

\bibitem{deepecg}
U.~B. Baloglu, M.~Talo, O.~Yildirim, R.~S. Tan, and U.~R. Acharya,
  ``Classification of myocardial infarction with multi-lead ecg signals and
  deep cnn,'' \emph{Pattern Recognition Letters}, vol. 122, pp. 23 -- 30, 2019.

\bibitem{deepliver}
A.~Das, U.~R. Acharya, S.~S. Panda, and S.~Sabut, ``Deep learning based liver
  cancer detection using watershed transform and gaussian mixture model
  techniques,'' \emph{Cognitive Systems Research}, vol.~54, pp. 165 -- 175,
  2019.

\bibitem{alexnet}
A.~Krizhevsky, I.~Sutskever, and G.~E. Hinton, ``Imagenet classification with
  deep convolutional neural networks,'' in \emph{Advances in Neural Information
  Processing Systems 25}, 2012, pp. 1097--1105.

\bibitem{deepface}
S.~{Lawrence}, C.~L. {Giles}, , and A.~D. {Back}, ``Face recognition: a
  convolutional neural-network approach,'' \emph{IEEE Transactions on Neural
  Networks}, vol.~8, no.~1, pp. 98--113, Jan 1997.

\bibitem{deepseg}
M.~{Kallenberg}, K.~{Petersen}, M.~{Nielsen}, A.~Y. {Ng}, P.~{Diao}, C.~{Igel},
  C.~M. {Vachon}, K.~{Holland}, R.~R. {Winkel}, N.~{Karssemeijer}, and
  M.~{Lillholm}, ``Unsupervised deep learning applied to breast density
  segmentation and mammographic risk scoring,'' \emph{IEEE Transactions on
  Medical Imaging}, vol.~35, no.~5, pp. 1322--1331, May 2016.

\bibitem{allan}
A.~Cerentini, D.~Welfer, M.~C. d'Ornellas, C.~J.~P. Haygert, and G.~N. Dotto,
  ``Automatic identification of glaucoma using deep learning methods,''
  \emph{MEDINFO 2017: Precision Healthcare through Informatics}, vol. 245, pp.
  318 -- 321, 2017.

\bibitem{cnn}
S.~{Lawrence}, C.~L. {Giles}, {Ah Chung Tsoi}, and A.~D. {Back}, ``Face
  recognition: a convolutional neural-network approach,'' \emph{IEEE
  Transactions on Neural Networks}, vol.~8, no.~1, pp. 98--113, Jan 1997.

\bibitem{translearn}
S.~J. {Pan} and Q.~{Yang}, ``A survey on transfer learning,'' \emph{IEEE
  Transactions on Knowledge and Data Engineering}, vol.~22, no.~10, pp.
  1345--1359, Oct 2010.

\bibitem{lbp2002}
T.~Ojala, M.~Pietikainen, and T.~Maenpaa, ``Multiresolution gray-scale and
  rotation invariant texture classification with local binary patterns,''
  \emph{IEEE Transactions on Pattern Analysis and Machine Intelligence},
  vol.~24, no.~7, pp. 971--987, Jul 2002.

\bibitem{lbpface}
T.~{Ahonen}, A.~{Hadid}, and M.~{Pietikainen}, ``Face description with local
  binary patterns: Application to face recognition,'' \emph{IEEE Transactions
  on Pattern Analysis and Machine Intelligence}, vol.~28, no.~12, pp.
  2037--2041, Dec 2006.

\bibitem{azar2014}
A.~T. Azar and S.~A. El-Said, ``Performance analysis of support vector machines
  classifiers in breast cancer mammography recognition,'' \emph{Neural
  Computing and Applications}, vol.~24, no.~5, pp. 1163--1177, 2014.

\bibitem{ura20171}
U.~Raghavendra, S.~V. Bhandary, A.~Gudigar, and U.~R. Acharya, ``Novel expert
  system for glaucoma identification using non-parametric spatial envelope
  energy spectrum with fundus images,'' \emph{Biocybernetics and Biomedical
  Engineering}, vol.~38, no.~1, pp. 170 -- 180, 2018.

\bibitem{bander}
B.~{Al-Bander}, W.~{Al-Nuaimy}, M.~A. {Al-Taee}, and Y.~{Zheng}, ``Automated
  glaucoma diagnosis using deep learning approach,'' in \emph{2017 14th
  International Multi-Conference on Systems, Signals Devices (SSD)}, March
  2017, pp. 207--210.

\bibitem{kirar1}
B.~S. {Kirar} and D.~K. {Agrawal}, ``Computer aided diagnosis of glaucoma using
  discrete and empirical wavelet transform from fundus images,'' \emph{IET
  Image Processing}, vol.~13, no.~1, pp. 73--82, 2019.

\bibitem{lbptext1}
Z.~{Guo}, L.~{Zhang}, and D.~{Zhang}, ``A completed modeling of local binary
  pattern operator for texture classification,'' \emph{IEEE Transactions on
  Image Processing}, vol.~19, no.~6, pp. 1657--1663, June 2010.

\bibitem{lbptext2}
S.~{Liao}, M.~W.~K. {Law}, and A.~C.~S. {Chung}, ``Dominant local binary
  patterns for texture classification,'' \emph{IEEE Transactions on Image
  Processing}, vol.~18, no.~5, pp. 1107--1118, May 2009.

\bibitem{tsne}
L.~{van der Maaten} and G.~Hinton, ``Visualizing high-dimensional data using
  t-sne,'' \emph{Journal of Machine Learning Research}, vol.~9, pp. 2579--2605,
  Nov 2008.

\end{thebibliography}

\end{document}